\documentclass[twocolumn,showpacs,showkeys,preprintnumbers,amsmath,amssymb,pra]{revtex4}

\usepackage{graphicx}
\usepackage{dcolumn}
\usepackage{bm}

\newcommand{\mcal}{\ensuremath{\mathcal}}
\newcommand{\opr}[1]{\ensuremath{\mathbf{\mathsf{#1}}}}

\newcommand{\expo}[1]{\ensuremath{\mbox{e}^{#1}}}
\newcommand{\abs}[1]{\ensuremath{\left|#1\right|}}

\newcommand{\dirac}[2]{\ensuremath{\left<#1\left|\right.#2\right>}}

\begin{document}

\preprint{xxx version}

\title{Transition from discrete to continuous time of arrival distribution\\
for a quantum particle}

\author{Eric A. Galapon$^{1,2,3}$}
\email{eric.galapon@upd.edu.ph}
\author{F. Delgado$^2$}
\author{J. Gonzalo Muga$^2$}
\author{I\~nigo Egusquiza$^3$}
\affiliation{$^1$Theoretical Physics Group, National Institute of Physics, University of
 the Philippines, Diliman, Quezon City, 1101 Philippines}
\affiliation{$^2$Departamento de Qu\'\i mica F\'\i sica, UPV-EHU, Apdo. 
644, 48080 Bilbao, Spain}
\affiliation{$^3$Theoretical Physics, The University of the Basque Country, Apdo. 
644, 48080 Bilbao, Spain}

\date{\today}

\begin{abstract}
We show that the Kijowski distribution for time of arrivals in the 
entire real line is the
limiting distribution of the 
time of arrival distribution in a
confining box as its length increases to infinity.
The dynamics of the confined time of arrival eigenfunctions is also 
numerically investigated and demonstrated that the eigenfunctions evolve 
to have point supports at the arrival point at their respective eigenvalues
in the limit of arbitrarilly large confining lengths, giving insight into
the ideal physical content of the Kijowsky distribution.
\end{abstract}

\maketitle

\section{Introduction}
The problem of accomodating time as a quantum dynamical observable has a long history and remains controversial to this day \cite{pauli,oth,povm,galapon,GCB04,time,gen,Kijowski74,All,Gia}.
The time of arrival (TOA) of a quantum structureless particle at a
point or a surface is one aspect of this problematic treatment of
time in quantum mechanics in need of clarification \cite{time,gen}.
Recently significant progress has been made, 
both at the operational and foundational fronts for the quantum time of arrival problem. 
Most notable at the operational level is
the convergence of several analyses of the problem to the axiomatic
distribution of arrival for freely moving particles due to Kijowski
\cite{Kijowski74}. Kijowski's distribution was first obtained
operationally 
(under some limiting conditions) by Allcock \cite{All} and 
later shown to be arising from the quantization of the
classical time-of-arrival \cite{Gia,MLP98}, inspiring
extension of the theory to the interacting case \cite{BSPME00,LJPU00,BEM01b} or
multiparticle systems \cite{BEM02}.  
Most important of these
developments in understanding the physical content of Kijowski
distribution is the realization that, while axiomatic in nature, it
can be obtained from an operational procedure
\cite{DEHM03,HSM03}, whose essence is to modify (filter) the initial 
state to counterbalance the disturbance introduced by the apparatus, 
in particular at low energies.   

At the foundational level, it has been shown that the 
non-self-adjointness of the free time of arrival operator, widely regarded as due
to the semiboundedness of the Hamiltonian in accordance with Pauli's 
theorem \cite{pauli,galapon}, can in fact be lifted by spatial confinement \cite{galapon}.
Thus the concept of confined quantum time of arrival (CTOA) was introduced
\cite{GCB04}. The CTOA-operators form a class of compact and self-adjoint
operators canonically conjugate with their respective Hamiltonians in a closed
subspace of the system Hilbert space. Being compact, the CTOA-operators
posses discrete spectrum, and a complete set of mutually orthogonal square
integrable eigenfunctions \footnote{An earlier work by Grot et. al. \cite{grot} addresses the non-self-adjointness of the quantized classical time of arrival in the entire real line by modifying the behavior of the TOA-operator in the neighborhood of vanishing momentum. Their operator is self-adjoint but fails to satisfy conjugacy with the Hamiltonian.}. The eigenfunctions of the CTOA-operators are found to be states that evolve to unitarily (i.e. according to Schrodinger's equation) 
arrive at the origin at their respective eigenvalues---that is the events
of the centroid of the position distribution being at the origin and its width
being minimum occur at the same instant of time. 

Now we are confronted with the problem of relating the already established results for the unconfined quantum particle to the confined one, in particular the question whether the established Kijowski time of arrival distribution is extractable or not from the confined time of arrival operators. Therefore, in this paper we give meaning to the limit of the discrete time of arrival distribution of the CTOA-operators defined on succesively larger segments, proving that this limit is the Kijowski's distribution. We begin by a short review of the properties of the confined and unconfined time of arrival operators; this is followed by numerical indications of the relation between these two cases, numerical indications that are strengthened by the analytical proof of the following statement: The time of arrival operator on the full real line is the limit for $l$ tending to infinity of the time of arrival operators defined for free motion in a segment of length $2l$. The dynamics of the CTOA-eigenfunctions are then numerically investigated in the limit of arbitrarilly large confining lengths. We end by summarising and providing conclusions.

\section{The Quantum free Time of arrival operator}
From a heuristic perspective, it is quite natural to assume that 
the quantum TOA may be associated with the quantization of the  
corresponding classical expression. 
That is, if a classical free particle, of mass $\mu$ 
in one dimension at initial location $q$ with momentum $p$, will arrive, say, at
the origin at the time $T(q,p)=-\mu qp^{-1}$, then the quantum TOA-distribution
must be derivable from a quantization of $T(q,p)$, such as the symmetrized 
\begin{equation}
\opr{T}=-\frac{1}{2}\mu (\opr{qp^{-1}+p^{-1}q}).\label{toa}
\end{equation}
Formally the time of arrival operator $\opr{T}$ is canonically
conjugate to the free Hamiltonian, $\opr{H}=(2\mu)^{-1}\opr{p}^2$,
i.e. $[\opr{H},\opr{T}]=i\hbar$.  Equation-(\ref{toa}) has been separately studied
when the motion of the free particle takes place in the entire
real line and when it is restricted to a segment of the real
line. These studies led to some seemingly contradictory descriptions, in
that the time of arrival operator is not self-adjoint for the full
real line, while there is an infinite family of self-adjoint operators
playing the corresponding role for the segment. Furthermore, for the
full line case there is a well defined property of time covariance
which is lacking in the confined case.

\subsection{The TOA-operator in $\mcal{H}_p=L^2(-\infty,\infty)$}
In momentum representation, Eq. (\ref{toa}) formally assumes the form
\[\opr{T}=\frac{i\hbar \mu}2\left(\frac1{p^2}-\frac2p
\frac{\partial}{\partial p}\right)\,.\]
Under a sensible choice of domain \cite{EM00}, $\opr{T}$ is a densely defined
operator, and unbounded in ${\cal H}_{p}:=L^{2}\left({\bf R},{\rm
d}p\right)$. It is however not self-adjoint, and admits no
self-adjoint extension. It is in fact a maximally symmetric
operator \cite{EM00}. According to the well-established theory for this kind of
operator, there exists a decomposition of unity given by the following
degenerate weak (or non-square integrable) eigenfunctions
\[\tilde\psi^{(t)}_{\alpha}(p)=\Theta(\alpha p)\left(\frac{\alpha
p}{2\pi \mu \hbar}\right)^{1/2} e^{i p^2 t/2 m\hbar}\,.\]
This is a family of functions parametrized by the eigenvalue $t$ and a
discrete parameter $\alpha$, which can be either $+1$ or $-1$, and
gives the sign of the half-line on which the function presents its
support in momentum space.  It is straightforward to prove
completeness, i.e.,
$\sum_{\alpha}\int_{-\infty}^\infty{\rm
d}t\,\tilde\psi_{\alpha}^{(t)}(p')\tilde\psi_{\alpha}^{(t)}(p)=\delta(p-p')$, 
and nonorthogonality, 
\[\int_{-\infty}^\infty{\rm d}p\,
\tilde\psi^{(t')}_{\alpha'}(p)\tilde\psi^{(t)}_{\alpha}(p)=\frac12
\delta_{\alpha\alpha'}\left(\delta(t-t')+\frac{i}{\pi}{\rm P}
\frac{1}{t-t'}\right)\,.\]
The physical content of the eigenfunctions is better examined in
coordinate representation \cite{MLP98}.  Normalized wave packets with
the form of quasi-eigenstates of $\opr{T}$ peaked at a given
eigenvalue, but with a time width $\Delta T$, have sharp space-time
behaviour. Their average position travels with constant velocity to
arrive at the origin at the nominal arrival time of the packet, in
which the packet also attains its minimum spatial width.  The passage
of probability density from one side of the origin to the other, is in
summary as sharp as desired by taking $\Delta T\to 0$ \cite{MLP98}.
As we shall see later, the unitary evolution of a (generalized) state
$|\tilde\psi^{(t)}_\alpha\rangle$ leads to a distribution with point
support at the origin at the instant corresponding to the eigenvalue
$t$, thus lending support to its interpretation as a time of arrival
eigenstate.

Using completeness, we can now write the 
probability density for measured values of the $\opr{T}$ operator,  
\begin{eqnarray}
\Pi_{\psi_0}(t) & = & \left|\int_{0}^\infty{\rm d}p\,\left(\frac{p}{2\pi
m\hbar}\right)^{1/2} e^{-ip^2 t/2m\hbar}\psi_0(p)\right|^2
+\nonumber\\ & &\quad
\left|\int_{-\infty}^0{\rm d}p\,\left(\frac{-p}{2\pi
m\hbar}\right)^{1/2} e^{-ip^2 t/2m\hbar}\psi_0(p)\right|^2\,,\nonumber
\end{eqnarray}
which is, in fact, Kijowski's probability density \cite{Kijowski74}.

An essential property of the distribution of arrivals in Kijowski's
axiomatic approach is covariance under transformations generated by
the Hamiltonian. It is evident that this property of covariance is
indeed held by the distribution above. Physically, it means that the
probability of arriving at $t$ for a given state is equal to the
probability of arriving at $t-\tau$ for the same state, once evolved a time
$\tau$.  This is the reflection on the probability density of the
canonical commutation relation between $\opr{H}$ and $\opr{T}$.

\subsection{The TOA-operator in $\mcal{H}_l=L^2[-l,l]$}
In \cite{GCB04} the naive definition in Eq. (\ref{toa}) 
$\opr{T}$ for a free particle in a segment of length $2l$ 
is supplemented by adequate boundary conditions, leading to
properly self-adjoint operators.
Physically, the choice of time of arrival operator with  spatial 
confinement of the particle in the
interval $[-l,l]$ is dictated by the condition that the evolution of the system is
generated by a purely kinetic self-adjoint Hamiltonian, i.e.
$\opr{H}=(2\mu)^{-1} \opr{p}^2$, where $\opr{p}$ is a self-adjoint momentum 
operator. This requirement demands that  the momentum operator
$\opr{p}$ be one of the set $\left\{\opr{p_{\gamma}}=-i\hbar \partial_{q},\,
\abs{\gamma}\leq\pi/2\right\}$, with $\opr{p}_{\gamma}$ having the  domain 
consisting of absolutely continuous functions $\phi(q)$ in
$\mcal{H}_l=[-l,l]$ with square integrable first derivatives, which
further satisfy the boundary condition
$\phi(-l)=\expo{-2i\gamma}\phi(l)$. Since $\opr{T}$ depends on the
momentum operator, $\opr{T}$ is also required to be the corresponding
element of the set of operators $\left\{\opr{T}_{\gamma}\right\}$.

In coordinate representation, $\opr{T}_{\gamma}$ becomes the Fredholm integral
operator $\left(\opr{T}_{\gamma}\varphi\right)\!(q)
=\int_{-l}^{l}T_{\gamma}(q,q')\,\varphi(q')\,dq
',$  for all $\varphi(q)$ in $\mcal{H}$, where the kernel is given by 
\begin{equation} \label{repre}
T_{\gamma\neq 0}(q,q')=-\mu\frac{(q+q')}{4\hbar\sin\gamma}\left(e^{i\gamma}
H(q-q')+e^{-i\gamma}H(q'-q) \right),
\end{equation}
\begin{equation} \label{periodic}
T_{\gamma=0}(q,q')=\frac{\mu}{4i \, \hbar}(q+q')\mbox{sgn}(q-q')-\frac{\mu}{4i\,\hbar
 l}\left(q^2 -q'^2\right),
\end{equation}
in which $H$ is Heaviside's step function and $\mbox{sgn}$ is the sign function.

With this representation, one can show that $\opr{H}_{\gamma}$ and
$\opr{T}_{\gamma}$ form a canonical pair in a closed subspace of
$\mcal{H}_l$---a non-dense subspace---for every $\gamma$. Moreover, the kernel $T_{\gamma}(q,q')$ of $\opr{T}_{\gamma}$ is
square integrable,
i.e. $\int_{-l}^l\int_{-l}^l\abs{T_{\gamma}(q,q')}^2\, dq dq'<\infty$.
This means that $\opr{T}_{\gamma}$ is compact, and, as a consequence,
that it has a complete set of (square integrable) eigenfunctions and
its spectrum is discrete.  This, it should be stressed, is a radically
different situation from that in the full line, where the operator has
a continuous spectrum and is not self-adjoint.

In what follows, we will need only the spectral properties for the
periodic confined quantum time of arrival operators (that is to say,
$\gamma=0$). The operator $\opr{T}_0$ 
commutes with the parity
operator. Furthermore, it changes sign under time inversion, which
entails that its spectrum is symmetric about 0. This suggests
classifying its eigenfunctions in even and odd subspaces (which will
be denoted by the subscripts $e$ and $o$ respectively), and, within
each of those subspaces, by a discrete index $n$ and the sign of the
eigenvalue (indicated as a superscript). In this manner, and as
computed elsewhere \cite{GCB04}, the odd eigenfunctions are
\begin{equation}
\varphi_{n,o}^{\pm}(q)=A_n q f^{\pm}\left(\frac{s_n q^2}{l^2}\right)
\label{eigenzerooddsimp}
\end{equation}
with
\begin{equation}
f^{\pm}(\xi)=e^{\mp i \xi} \xi^{1/4}\left[J_{-\frac{1}{4}}(\xi) \mp i J_{\frac{3}{4}}(\xi)\right]
\label{defnf}
\end{equation}
while the even ones are given by
\begin{equation}
\varphi_{n,e}^{\pm}(q)=B_n g^{\pm}\left(\frac{r_n q^2}{l^2}\right)
\end{equation}
with
\begin{equation}
g^{\pm}(\xi)=e^{\mp i \xi} \xi^{3/4}\left[J_{-\frac{3}{4}}(\xi)\mp i J_{\frac{1}{4}}(\xi)\right]+\frac{e^{\mp i r_n} J_{\frac{1}{4}}(r_n)}{r_n^{\frac{1}{4}}}\,.
\label{defng}
\end{equation}
$A_n$ and $B_n$ are the normalization constants, while $s_n$ and $r_n$ are the solutions of the secular equations for the operator, namely the positive roots of  $J_{-\frac{3}{4}}(r)
+\frac{2}{3}J_{\frac{5}{4}}(r)+\frac{1}{r}J_{\frac{1}{4}}(r)=0$ for the even case and of $J_{-1/4}(s)=0$ for the odd case. The eigenvalues are determined by  
\begin{equation}
\tau_{n}^{\pm}=\pm \frac{\mu l^{2}}{4\rho_{n}\hbar}\,,
\label{eigeneven}
\end{equation}
where $\rho_n$ stands for either $r_n$ or $s_n$.

It has been demonstrated in \cite{GCB04} that a CTOA eigenfunction is a state
that evolves unitarily---that is according to Schrodinger's equation---to being concentrated around the origin at its
eigenvalue
along a classical trajectory, i.e. a state in which the events of the centroid
of the position distribution being at the origin and its width being minimum occur
at the same instant of time equal to the eigenvalue. 

\section{Recovering Kijowski's distribution from the CTOA-operators}

\subsection{Numerical examples}

We have thus seen that the time of arrival operators for confined and
unconfined free motion have radically different properties. And it is not immediate how Kijowski's distribution follows from the spectral properties of the CTOA-operators.
In this section we show that indeed Kijowski's distribution is extractable from the discrete time of arrival distribution defined by the CTOA-operator. But before doing so we first provide
further justification from numerical results.

One basic difference between the two cases being examined is that the
spectrum is continuous for the full line, whereas it is discrete for
the segment. As a consequence, the same can be predicated of the
probability distributions for times of arrivals. Therefore, in order
to compare like with like, it is convenient to use the accumulated
probability of having arrived at the origin before a given instant for
both cases.

Also in order to perform sensible comparisons we shall consider
initial states with compact support on the real line, which can also
be defined for all segments which would include that compact
support. In this manner the probability of having arrived at the
origin before instant $t$ can be computed for the same function both
for the segment and the line.

In the case of the segment, the probability of having arrived at the
origin before instant $t$ for the initial state $\psi_0$ is calculated
according to the standard quantum mechanical prescription,
\begin{equation}\label{accud}
F_{\psi_0}(t)=\sum_{\tau_{\gamma,s}\leq t}\abs{\dirac{\varphi_{\gamma,s}}{\psi_0}}^2,
\end{equation}
where $\varphi_{\gamma,s}$ and $\tau_{\gamma,s}$ are the eigenfunctions 
and eigenvalues of $\opr{T}_{\gamma}$. This distribution is in turn 
investigated for increasing $l$, and compared with the accumulated 
probability for Kijowski's distribution,
\begin{equation}
F^K_{\psi_0}(t)=\int_{-\infty}^t\Pi_K\left[\tau,\psi_0\right]\, d\tau.
\end{equation}
where $\Pi_K\left[\tau,\psi_0\right]$ is the Kijowsky time of arrival density.

As the length of the segment $l$ increases, the discrete eigenvalues
of the corresponding time of arrival operators become denser, and we
take advantage of this property to have another representation for
comparison with Kijowski's distribution; namely plotting Kijowski's
time of arrival density together with a discrete derivative of the
probability of having arrived at the origin before a given instant, in
the case of the segment, discrete derivative that can be understood as
a time of arrival density. This allows us an easier recognition of
some features that would otherwise be obscured in the accumulated
probability. 

\begin{figure}[!tbp]
{\includegraphics[height=3.25in,width=2.0in,angle=270]{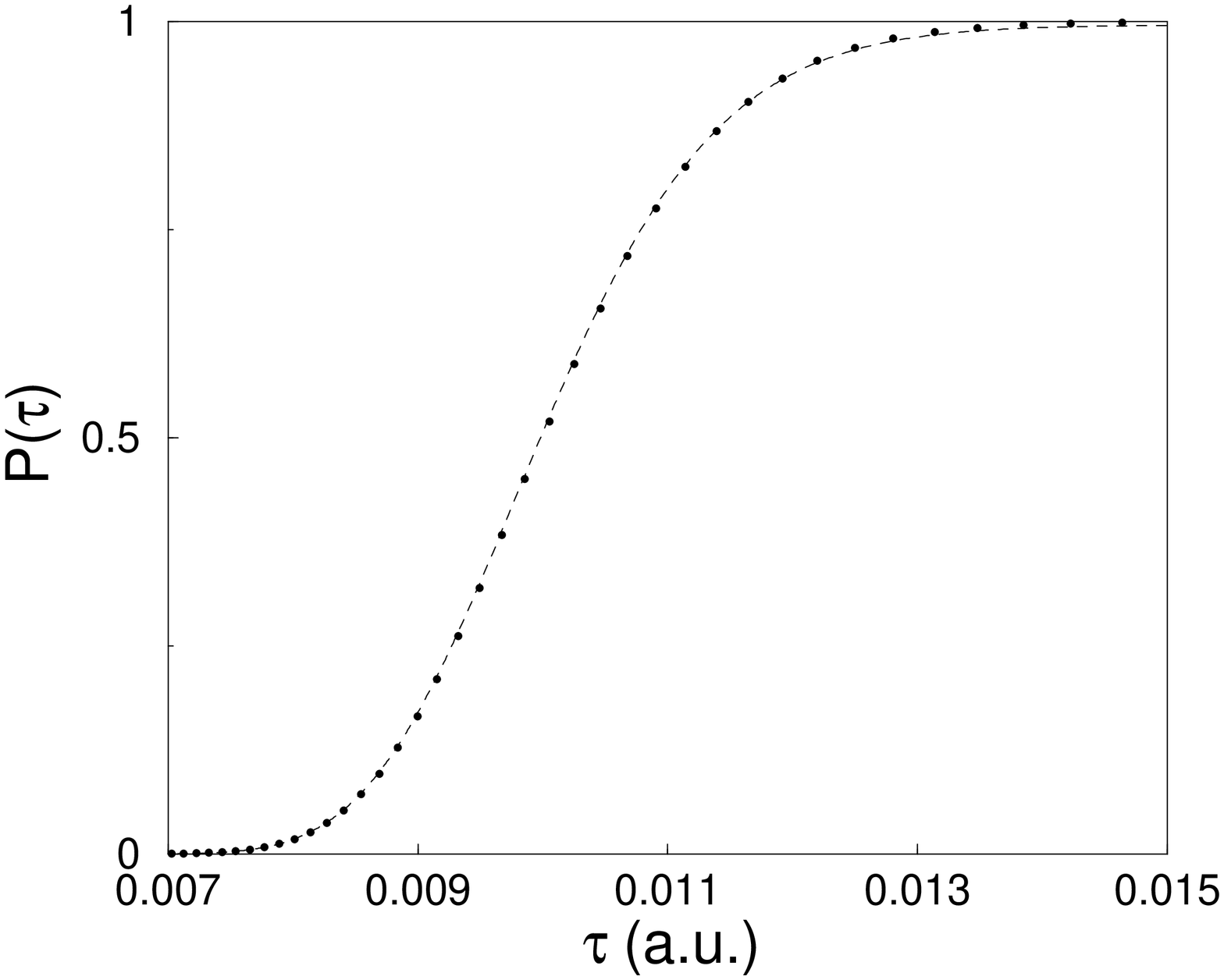}}
{\includegraphics[height=3.25in,width=2.0in,angle=270]{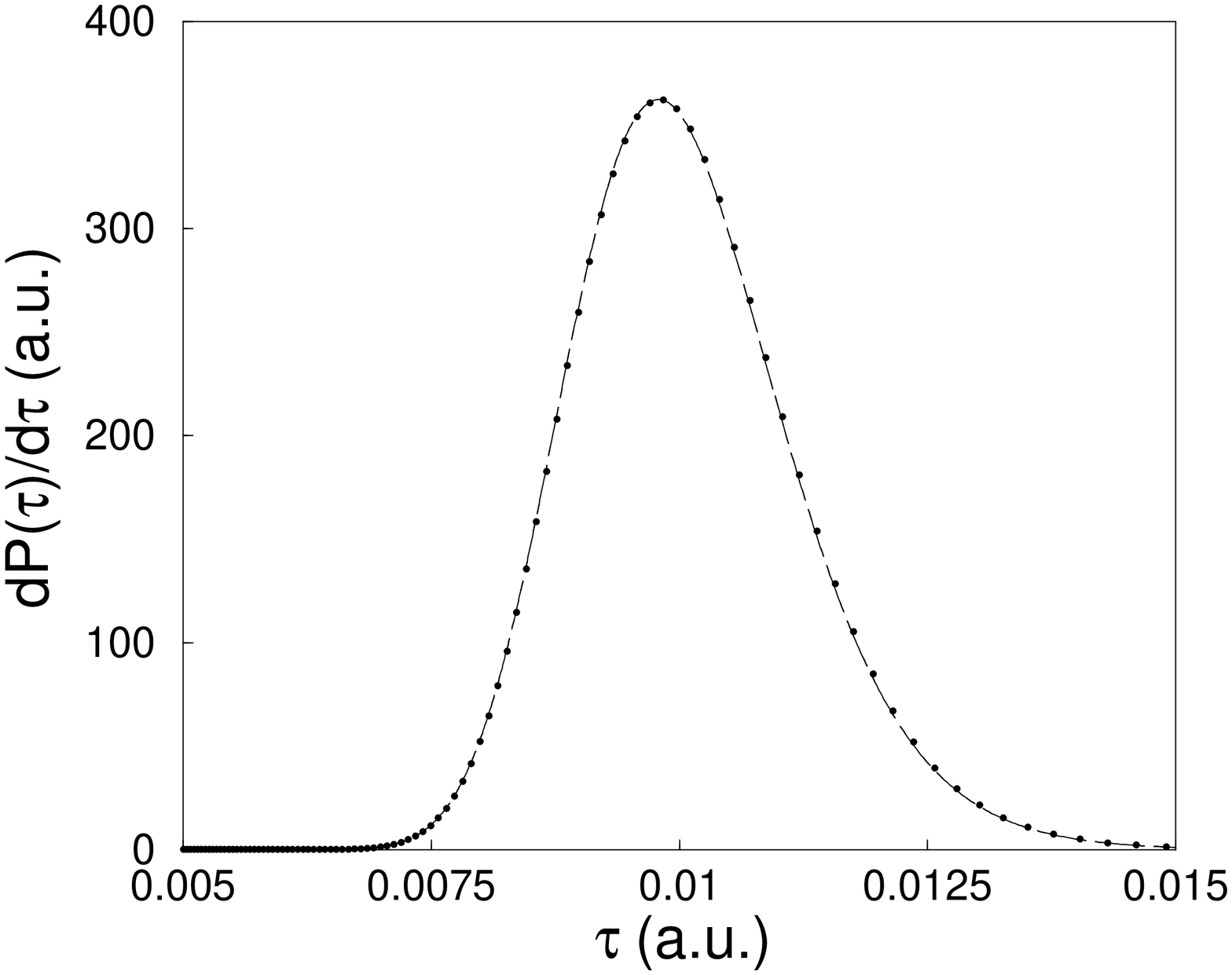}}
\caption{The top-figure shows the accumulated probability of arrival versus
time  at the origin for a gaussian wave packet
with mean momentum $<P>=100$, 
full width at half maximum $\sigma_x=0.05$ and initial expected value
 $<X>=-1$ (all the
quantities in atomic units). The dots correspond to confined motion with $l=10$, and the solid line to the full line.
In the lower figure we depict the corresponding probability densities 
(with the same notation), defined as explained in the text.}
\label{fig1}
\end{figure}

\begin{figure}[!tbp]
{\includegraphics[height=3.25in,width=2.0in,angle=270]{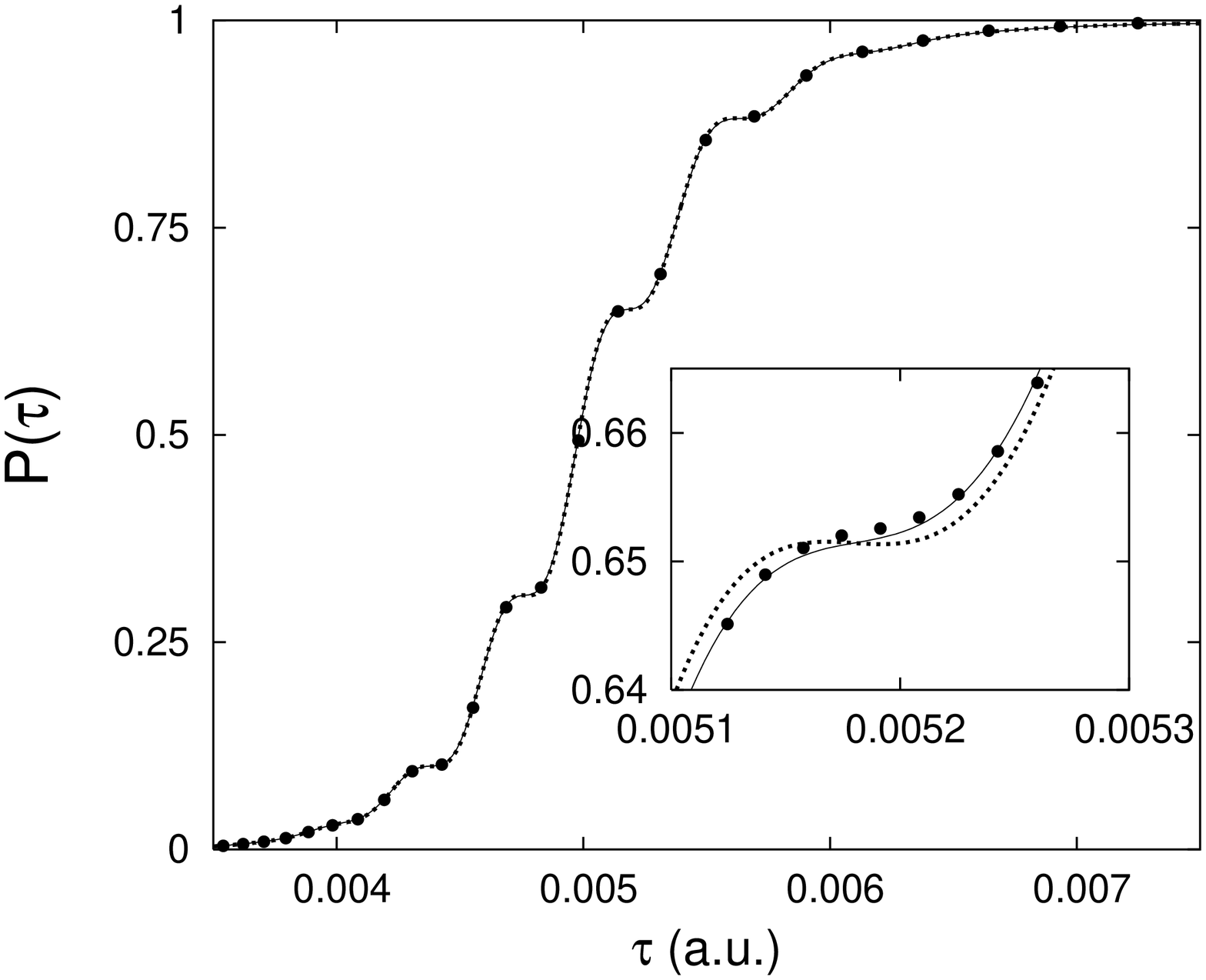}}
{\includegraphics[height=3.25in,width=2.0in,angle=270]{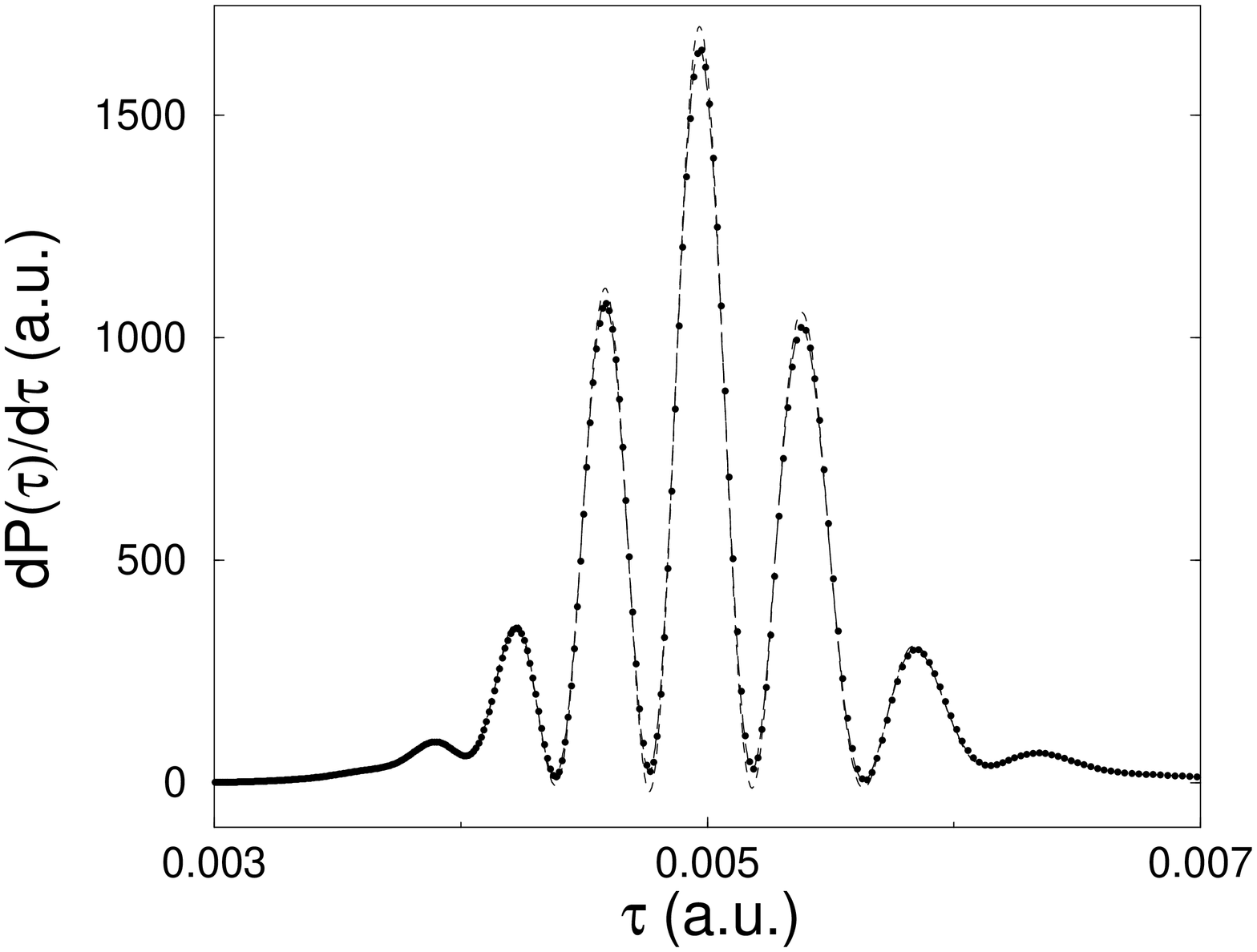}}
\caption{Accumulated probability of arrival (upper figure), versus time 
at the origin for two gaussian 
wave packet with mean momentum $<P_1>=200$ and $<P_2>=100$
with maximum interference at the origin, e.g., $<X_1>=-1$
and $<X_2>=-0.5$. The full width at half maximum is $\sigma_x=0.05$
for both gaussians.
The dots correspond to confined motion with $l=10$, while the solid line corresponds to the full line. For completeness the dotted line depicts the quantum probability flux at the origin, integrated up to the relevant instant. The inset shows the zone of maximal discrepancy between integrated flux and accumulated Kijowski's probability, where the accumulated probability for confined motion matches Kijowski's.
In the lower figure we show the corresponding probability densities 
(with the same notation) defined as explained in the text, as well as the quantum probability flux at the origin: in this situation there is a
backflow effect \cite{BM94}.} 

\label{fig2}
\end{figure}

\begin{figure}[!tbp]
{\includegraphics[height=3.25in,width=2.0in,angle=270]{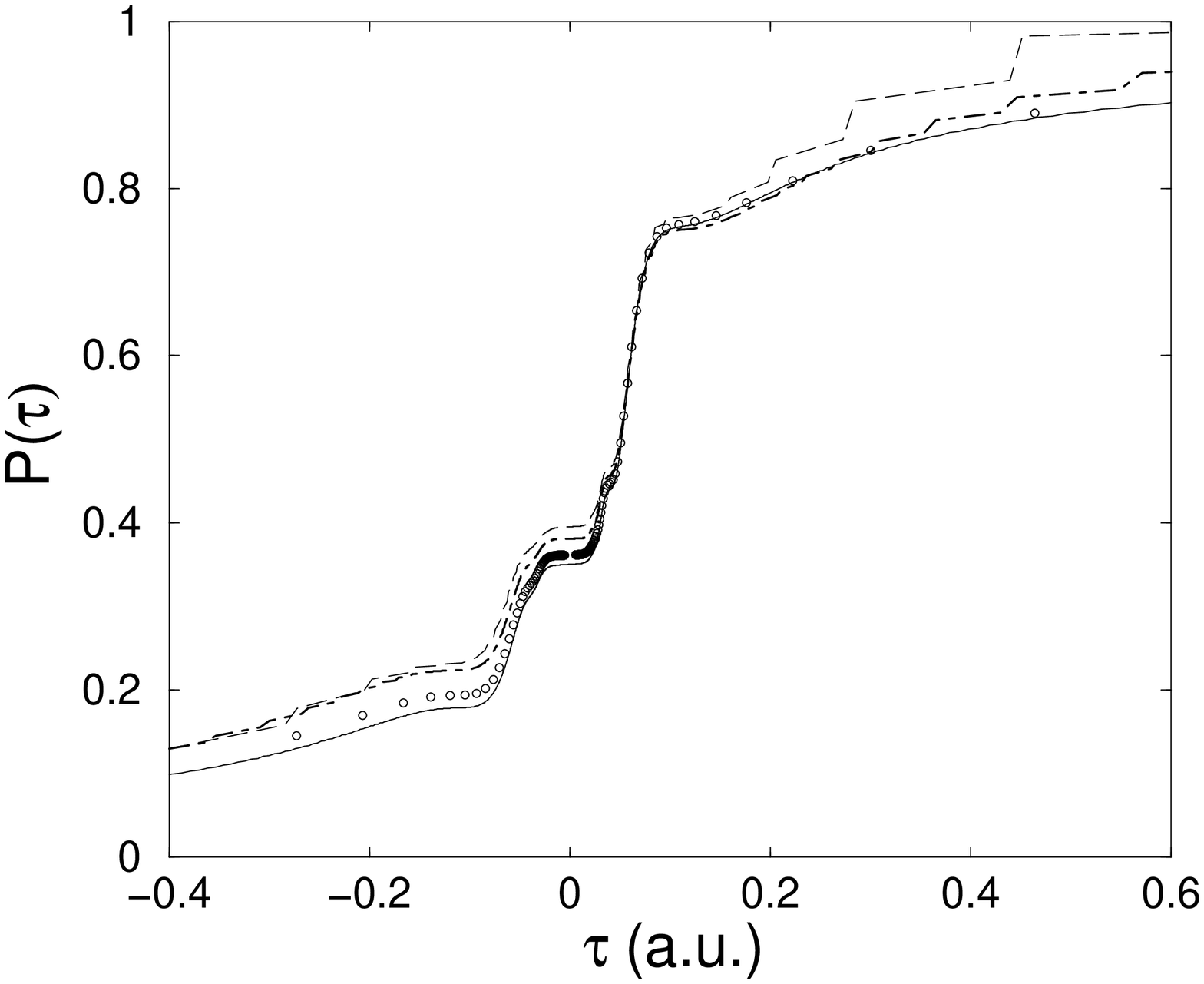}}
{\includegraphics[height=3.25in,width=2.0in,angle=270]{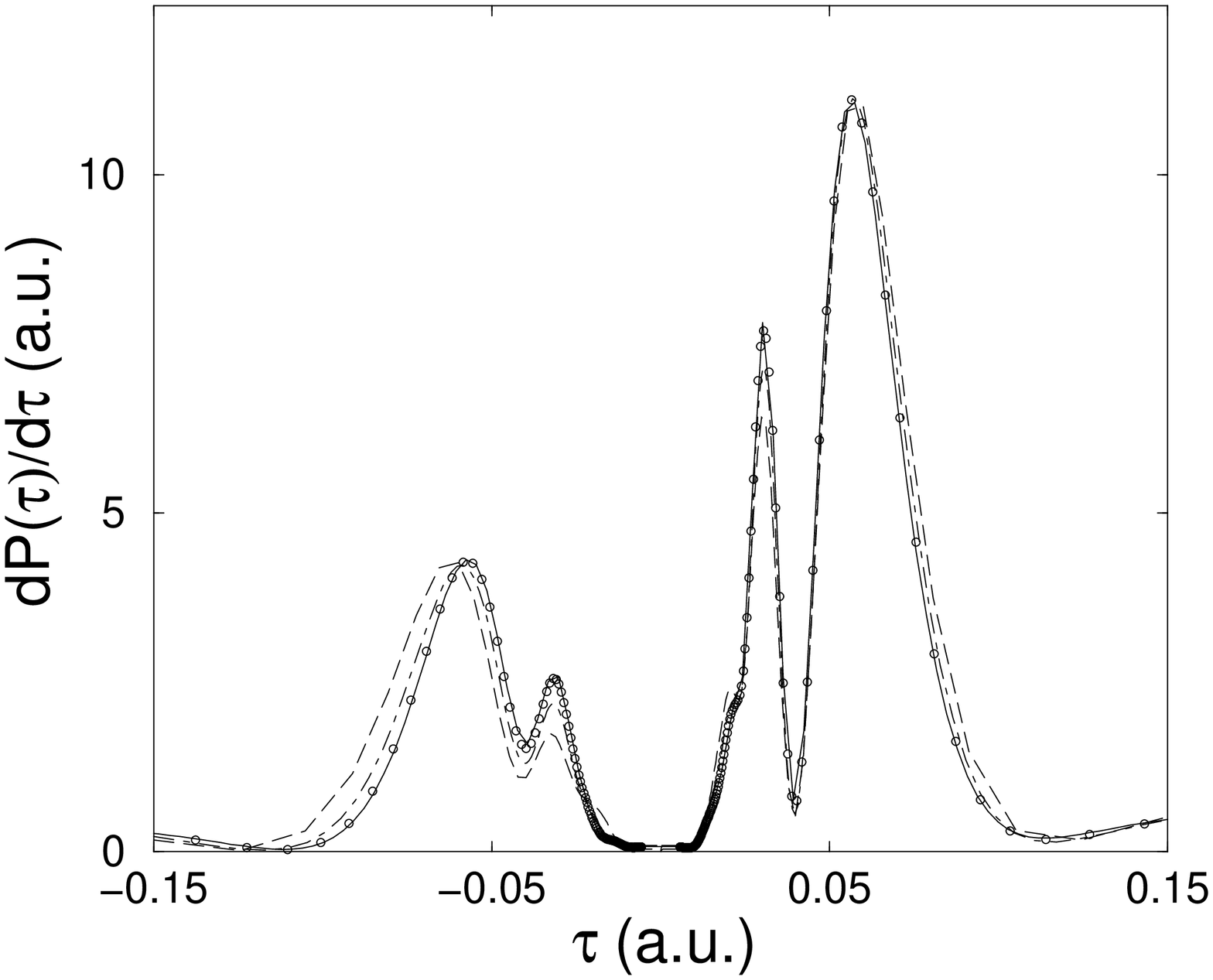}}
\caption{The upper figure shows the accumulated probability of arrival at the origin versus
time for two gaussian wave packets with mean momentum $<P_1>=5$ and
$<P_2>=1.5$ with $<X_1>=-1$ and $<X_2>=-0.5$. The full width at half
maximum is $\sigma_x=0.05$ for both gaussians. The distinct lines
correspond to different values of $l$: $l=3$, long dashed line; $l=5$,
dotted-dashed line, and $l=15$, circles. The solid line corresponds to
Kijowski's distribution.  In the lower figure we show the
corresponding probability densities (with the same notation), as
defined in the text.}
\label{fig3}
\end{figure}

Figs. \ref{fig1}, \ref{fig2} and \ref{fig3} show the accumulated probability
for the discrete
case and for the Kijowski distribution, together with their
corresponding time of arrival densities. In Fig. \ref{fig1} we depict a simple situation for an initial
gaussian state, with good match between the distributions for both the
full line and the segment. Fig. \ref{fig2} corresponds to an initial
state that presents the backflow effect \cite{BM94}, that is, that at
some instants the quantum probability flux can become negative even if
all the components of the state are of positive momentum. In the
situation depicted in Fig. \ref{fig2}, therefore, we can discriminate
whether the discrete distribution associated with the segment
approaches Kijowski's distribution or, rather, the flux, which is also
related to the density of arrivals. The result, as can be ascertained
from the inset in Fig. \ref{fig2}, is that Kijowksi's distribution is
the one selected in the limit of large $l$. Fig. \ref{fig3} is a depiction of accumulated probabilities for
different values of $l$, clearly showing convergence to Kijowski's
accumulated probabilities; similarly with the discrete derivatives.

All together, these numerical results suggest strongly that the
$l\to\infty$ limit of the time of arrival operators for the segment of
length $2l$ tend to the time of arrival operator on the real line, in
some sense.

\subsection{The limit for large $l$}
A first hint into the relationship between the confined and unconfined TAO-operators
is provided by the kernel of the the later in position representation.
In $q$-representation, the $\opr{T}$
operator takes on the form of the integral operator
\begin{equation}
\left(\opr{T}\varphi\right)\!(q)=
\int_{-\infty}^{\infty}\left<q\right|\opr{T}\left|q'\right>\varphi(q')\, dq'
\end{equation}
where the kernel is given by
\begin{equation}
\left<q\right|\opr{T}\left|q'\right>=\frac{\mu}{4 i \hbar}
\left(q+q'\right)\mbox{sgn}(q-q').
\end{equation}
Clearly this is the $l\to\infty$ formal limit of the kernel $T_0(q,q')$.
It is also the limit as $\gamma\to0$ of kernels $T_\gamma(q,q')$,
and, since the large $l$ limit washes out the effect of the boundary conditions, (i.e. the wavefunctions must necessarilly vanish at infinity)
this implies that the kernels match up in the large $l$ limit. For this reason, it will be sufficient for us to consider the periodic confined time of arrivals in the limit of arbitrarilly large confinement lengths.
\subsubsection{Kijowski distribution in position representation}
In what follows we will need the position representation of
the degenerate eigenfunctions of $\opr{T}$. In $q$-representation, 
the eigenfunctions assume the form
\begin{equation}\label{pospos}
\psi_t^{\pm}(q)=\frac{1}{\sqrt{2\pi \hbar}}\int_0^{\infty} 
e^{\pm iq p/\hbar} \left(\frac{p}{2\mu\hbar}\right)^{\frac{1}{2}}
e^{i p^2 t/2\mu\hbar} \, dp,
\end{equation}
where the integration is understood in the distributional sense.
Even and odd combinations of these are likewise eigenfunctions with the same
eigenvalue $t$,
\begin{equation}
\varphi_t^{e/o}(q)=\frac{1}{\sqrt{2}}\left(\psi_t^{+}(q) \pm \psi_t^{-}(q)\right)
\end{equation}
%
The full explicit expression for $\varphi_t^{e/o}(q)$ is obtained by direct substitution, leading to 
\begin{eqnarray}
\varphi_t^{e/o}(q)
&=& e^{i\frac{(1\pm 2)\pi}{8}}\frac{1}{4}\sqrt{\frac{2}{\pi t}} \left(\frac{\mu}{\hbar t}\right)^{\frac{1}{4}} \exp\left({-i \frac{\mu q^2}{4\hbar t}}\right)\nonumber\\
& &\!\!\!\!\!\!\!\! \left[D_{\frac{1}{2}}\!\left(-e^{i\frac{\pi}{4}}\sqrt{\frac{\mu}{\hbar t}}q\right) \pm D_{\frac{1}{2}}\!\left(e^{i\frac{\pi}{4}}\sqrt{\frac{\mu}{\hbar t}}q\right) \right]\label{rere}
\end{eqnarray}
where $D_{\nu}$ is the parabolic cylinder function, and the even and
odd case correspond to positive and negative sign respectively.

In momentum representation it is known that these
eigenstates form a resolution of the identity. It thus follows, owing from the unitarity of the Fourier transform, that
the same happens in the position representation Hilbert space,
$L^2(\mathbb{R})$, i.e. $\sum_{\beta}\int_{-\infty}^\infty
dt\,\overline{\varphi^{\beta}_{t}}(q')\varphi^{\beta}_{t}(q)
=\delta(q-q') $, with $\beta$ standing for $e$ and $o$. They are
likewise non-orthogonal.

Because of the invariance of Kijowski's distribution under parity,  it can be
split in even and odd components of the wavefunction. In terms of $\varphi_t^{e/o}(q)$, it then
assumes the form
\begin{equation}
\Pi_{\psi_0}(t)=\Pi_{\psi_0}^e(t)+\Pi_{\psi_0}^o(t)\label{kiwpos}
\end{equation}
where
\begin{equation}
\Pi_{\psi_0}^{e/o}(t)=\abs{\int_{-\infty}^{\infty}
\overline{\varphi_t^{e/o}(q)}\psi_0(q)\, dq}^2\,.
\end{equation}
In the following section, we will show that $\Pi_{\psi_0}^{e/o}(t)$, and, hence,
Kijowski's distribution itself, can be obtained from the accumulated probability
of arrival $F_{\psi_0}^{(l)}\!(t)$ in the limit $l\to\infty$.

We will do so by a proper identification of the limit of the eigenfunctions 
of the confined quantum time of arrivals for $l$ approaching infinity as the
eigenfunctions given
by Eqs. (\ref{rere}). To this end we will need the position representation of
the eigenvalue problem for $\opr{T}$, which is explicitly given in momentum
representation in the form
\begin{equation}
\frac{i\mu\hbar}{2}\frac{1}{p^2} \psi_{t}(p)-i\mu\hbar
\frac{1}{p}\frac{d \psi_{t}(p)}{dp}=t \psi_{t}(p).
\end{equation}
Multiplying both sides of the equation by $p^2$ and then Fourier transforming
the resulting expression leads to the position representation of the above
eigenvalue equation
\begin{equation}
\frac{d^{2}\varphi_{t}(q)}{dq^{2}}+\frac{\mu iq}{t \hbar}\frac{d \varphi_{t}(q)}{dq}+\frac{3\mu i}{2t \hbar}\varphi_{t}(q)=0
\label{infl}
\end{equation}
where 
$\varphi_{t}(q)=\frac{1}{\sqrt{2\pi \hbar}}
\int_{-\infty}^{\infty}\exp\left(\frac{i}{\hbar}q\, p \right)\psi_{t}(p) \, dp$.
Straightforward substitution of $\varphi_t^{e/0}(q)$ in the differential
equation shows that they are linearly independent solutions of
Eq. (\ref{infl}).

\subsubsection{$F_{\psi_0}^{(l)}(t)$ for large $l$}

As we have pointed out above, both in the case of the full real line
and in that of the segment the Hamiltonian and the operators $\opr{T}$
and $\opr{T}_0$ are invariant under parity. It follows that, if we
rewrite an initial function in terms of an even and an odd components,
their separate distributions for times of arrival sum to the
distribution for the total function, with no interference term being
required. It is thus useful to separate the analysis in the even and
odd sectors. That is, we need to examine whether the distributions
$F_{\psi_0}^{(l) e/o}(t)$ (which are the separate contributions
to the probability of having arrived at the origin prior to instant
$t$ of the even and odd components of the initial state $\psi_0$)
do indeed tend to
$\int_{-\infty}^t\mathrm{d}t'\,\Pi_{\psi_0}^{e/o}(t')$ respectively as
$l$ approaches infinity.

First let us consider the contributions from the odd eigenfunctions for the
accumulated probability. This is given by
\begin{widetext}
\begin{equation}
F_{\psi_0}^{(l)o}(t)=
\sum_{\tau_{n}^{\pm}\leq t} \frac{1}{l^3 J_{\frac{3}{4}}^2(s_n) s_n^{1/2}}
\abs{\int_{-l}^l q e^{\pm i s_n q^2/l^2}
\left(s_n\frac{q^2}{l^2}\right)^{\frac{1}{4}} \left[J_{-\frac{1}{4}}
\left(s_n \frac{q^2}{l^2}\right) \pm i 
J_{\frac{3}{4}}\left(s_n\frac{q^2}{l^2}\right)\right] \,
\psi_0(q)\, dq}^2 \label{cow}
\end{equation}
\end{widetext}
where the sum includes only contributions from eigenfunctions with
eigenvalues smaller than $t$, $l$ is large enough such that the
interval $[-l,l]$ contains the support of $\psi_0(q)$, and we have
used the explicit value of the normalization factor $A_n$. 

As our numerical computations demonstrate, for increasing $l$  the
dominant eigenfunctions contributing in the calculation of the
accumulated probability come from large values of $n$. But as $n$
increases the eigenvalues of the dominant contributors become
denser. Therefore, for every time $t$ and large $l$ there exists a
corresponding $n(l,t)$ such that $|t|-\tau_{n(l,t)}$ (that is, the
difference between the eigenvalue closest to $|t|$ -- for that value
of $l$ -- and $|t|$) tends to 0 as $l$ tends to infinity. For any given $\tau$ then we can write the corresponding $s_n$ as as
$\mu l^2/4\tau\hbar$, with the adequate sign, in the inner
integrand. 

As for the normalization factor, it is clearly the case that $s_n$
tends to infinity, in which limit we can use the asymptotic properties
of Bessel functions and their zeroes to write $l^3
J_{\frac{3}{4}}(s_n)^2 s_n^{1/2}$ in the form $2 l^3/\pi s_n^{1/2}$
or, alternatively, $(4l^2/\pi)\sqrt{\hbar|\tau|/\mu}$.
Finally, since the spacing between the roots of $J_{-1/4}$ tends to
$\pi$ as $n$ tends to infinity, we have the result that the spacing
between successive values of $\tau_n$ (with either sign) is
$4\pi\hbar\tau^2/(\mu l^2)$, whence it follows that the sum
$\sum_{\tau_n^\pm\leq t}$ can be substituted by the integral
$\mu l^2\int_{-\infty}^t d\tau/4\hbar\pi\tau^2$

Putting together all these asymptotic expressions, we see that 
Eq. (\ref{cow}) becomes
\begin{widetext}
\begin{equation}
F_{\psi_0}^{(l)o}(t)\longrightarrow \frac{2\hbar}{\mu}\int_{-\infty}^{t}\,
d\tau\, \left(\frac{\mu}{4\tau \hbar}\right)^{\frac{5}{2}}
 \abs{\int_{-\infty}^{\infty} q e^{i \mu q^2/4\tau\hbar}
\left(\frac{\mu}{4\tau\hbar}q^2\right)^{\frac{1}{4}} \left[J_{-\frac{1}{4}}\left(\frac{\mu}{4\tau\hbar}q^2\right) + i J_{\frac{3}{4}}\left(\frac{\mu}{4\tau\hbar}q^2\right)\right] \, \psi_0(q)\, dq}^2 \label{cowcow}.
\end{equation}
Differentiating this with respect to time yields the corresponding probability 
density
\begin{equation}
\Pi_{\psi_0}^{odd}(t)=\frac{2\hbar}{\mu} \left(\frac{\mu}{4 \hbar t}\right)^{\frac{5}{2}}
 \abs{\int_{-\infty}^{\infty} q e^{i \mu q^2/4 t\hbar} \left(\frac{\mu}{4 t\hbar}q^2\right)^{\frac{1}{4}} \left[J_{-\frac{1}{4}}\left(\frac{\mu}{4 t\hbar}q^2\right) + i J_{\frac{3}{4}}\left(\frac{\mu}{4 t\hbar}q^2\right)\right] \, \psi_0(q)\, dq}^2 \label{cowc}.
\end{equation}
\end{widetext}
from which we extract the limit of the odd eigenfunctions as $l$ approaches
infinity. The limit is
\begin{eqnarray}
\varphi_t^{odd}(q)&=&\sqrt{\frac{2\hbar}{\mu}} \left(\frac{\mu}{4 \hbar t}\right)^{\frac{5}{4}} 
 q \left(\frac{\mu}{4 t\hbar}q^2\right)^{\frac{1}{4}} \exp\left({-i \frac{\mu q^2}{4 t\hbar}}\right)\nonumber\\  & & \times\left[J_{-\frac{1}{4}}\left(\frac{\mu}{4 t\hbar}q^2\right) - i J_{\frac{3}{4}}\left(\frac{\mu}{4 t\hbar}q^2\right)\right]\label{odd}
\end{eqnarray}
The relationship between equations-\ref{rere} and-\ref{odd} is not immediate. But their relationship can be established by substituting $\varphi_t^{odd}(q)$ back in equation-(\ref{rere}), and finding that it is a solution to the differential equation, meaning it is an eigenfunction of the time of arrival operator in the entire real line with the eigenvalue $t$.  Since $\varphi_t^{odd}(q)$ is odd, then it must differ at most with $\varphi_t^{-}(q)$ by a constant factor. Expanding $\varphi_t^{odd}(q)$ and $\varphi_t^{-}(q)$ about $q=0$, we find that they differ only only up to the irrelevant phase factor $e^{-i\pi/4}$. Then we must have $\Pi_{\psi_0}^{odd}(t)=\Pi_{\psi_0}^-(t)$ in the limit of infinite $l$.

The same procedure can be carried out for the contribution of the even
eigenfunctions, albeit with more cumbersome algebra, and the
corresponding conclusion is derived. As a consequence, we have proved
that the $l\to\infty$ limit of the discrete probability distribution
for times of arrival for a spatially confined particle is indeed Kijowski's
probability distribution.

It might be thought that our proof is lacking in that above we assumed
that the initial state has compact support, while Kijowski's
distribution must also be defined for other normalizable
functions. However, our analysis can be extended to initial states
with tails extending to infinity.  Let $\psi(q)$ be such a state. There
always exists a sequence of states  with compact
supports, $\psi_n(q)$, $n=1, 2,\dots$, such that $\psi_n(q)\rightarrow \psi(q)$. We can, say, pick
$\psi_1(q)$ and apply our above analysis to this initial state.  Once we get
the limit of infinite $l$, we follow it with the limit in $n$.  The
resulting limit is going to be Kijowski's distribution because
of the continuity of the inner product.

\section{The dynamics of the CTOA-eigenfunctions in the limit of infinite $l$}
We have shown above that Kijowski distribution is the limit of
the discrete distribution for arbitrarily large $l$. 
But what physical insight can we get from this realization?
Recall that one of the surrounding issues against an ideal
quantum time of arrival distribution is the fact that a quantum particle losses
the localized property of its corresponding classical particle entity.
Classically the concept of time of arrival is well-defined because a classical
particle has a well defined trajectory. This is contrary to the fact that no
such trajectory can be ascribed to the quantum particle---it has no definite position and momentum.

Now the theory of confined quantum time of arrivals demonstrates that the
quantum time of arrival problem, at the ideal level, i.e. at the level
where measuring instruments play no 
explicit role, can be rephrased to 
finding states that unitarily arrive at a given point at a definite 
time---states in which the events of the centroid being at the origin and
the position distribution width being minimum occur at the same instant of 
time. The QTOA-problem phrased in this way is  well-defined because
quantum states have well-defined trajectories according to the 
Schrodinger equation. All these give us insight to the ideal physical content of Kijowski
distribution. Let us see how. 

Let us consider some fixed time $\tau$, and for any given length $l$ of spatial 
confinement, we can find an $n$ such that the eigenvalue $\tau_n$ is
closest to $\tau$.
Now consider a sequence of monotonically increasing $l$'s, $l_1, l_2, l_3, ...$, 
with $l_1<l_2<l_3<...$. Then there will be an $n_1$ corresponding to 
$l_1$ such that $\tau_{n_1}$ is closest to $\tau$; and an $n_2$ corresponding
to $l_2$ such that $\tau_{n_2}$ is closest to $\tau$; and so on.
For arbitrarily large lengths, we should have 
$\tau_{n_1} \approx \tau_{n_2} \approx .... \approx \tau$,
so that in the limit of infinite lengths they converge to $\tau$.

We know that the eigenfunctions $\varphi_{n_1}, \varphi_{n_2}, ...$ 
will obtain their minimum variances at their respective eigenvalues
$\tau_{n_1}, \tau_{n_2}, ...$. How do the variances compare for the different 
$\varphi_n$'s? Equivalently, what is the behavior of the variance with
respect to $\varphi_n$ at the eigenvalue $\tau_n$ as $l$ approaches infinity?
In this case the variance increases with $l$. This is not surprising 
because the $\varphi_n$'s would be thrown out of the Hilbert space for
infinite $l$, i.e. they acquire infinite variances.

\begin{figure}[!tbp]
{\includegraphics[height=3.25in,width=2.0in,angle=270]{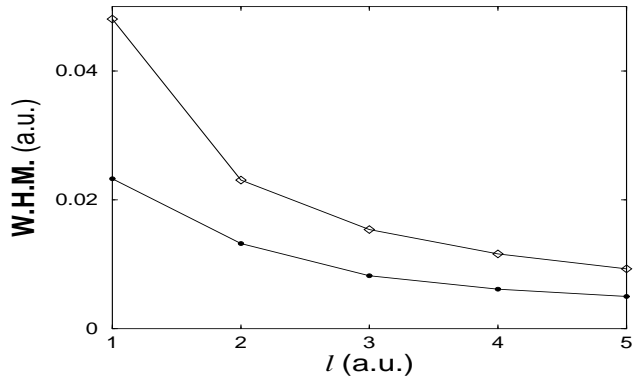}}
\caption{Width at half maximum (WHM) of the evolved 
odd and even eigenfunctions 
(upper and lower lines) at the corresponding eigenvalues closest to $t=0.01$
versus length $l$.}
\label{fig3:mwm}
\end{figure}

\begin{figure}[!tbp]
{\includegraphics[height=3.25in,width=2.0in,angle=270]{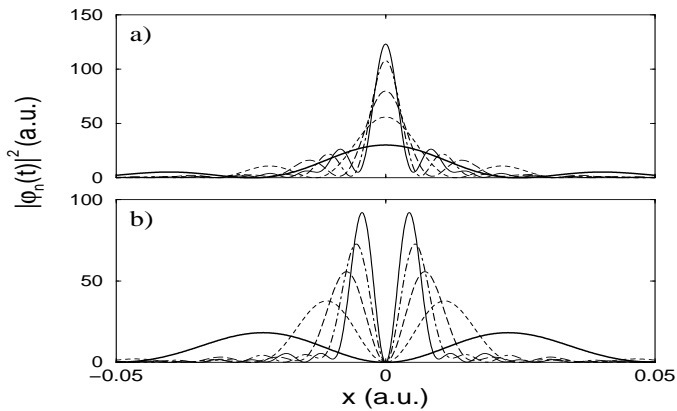}}
\caption{Probability density $|\varphi_n(\tau)|^2$ versus position at the 
corresponding closest eigenvalue $\tau_n$ to $t=0.01$, for the even 
(upper figure) and odd (lower figure) eigenfunctions. The different lines
are associated with $l=1$ (thick solid line), $l=2$ (dashed line), $l=3$
(long-dashed line), $l=4$ (dotted-dashedline) and $l=5$ (thin solid line).}
\label{fig2:mwm}
\end{figure}

However, if we substitute [width at half maximum (WHM) of the
probability density $\abs{\varphi_n(q,\tau_{n})}^2$ at
the eigenvalue $\tau_n$] for the variance,
we find that WHM decreases with increasing $l$.
Figure-\ref{fig3:mwm} demonstrate this. What is even more important 
is that the density $\abs{\varphi_{n}(q,\tau_n)}^2$ at the 
eigenvalue $\tau_n$ tends, as $l$ increases indefinitely, to a function with point support at the origin.
Figure-\ref{fig2:mwm} demonstrates this.
These results imply that the CTOA-eigenfunctions in the limit of 
infinite $l$ (which are already outside of the Hilbert space) evolve 
to ``collapse'' at the origin at their respective eigenvalues.
This allows us to interpret Kijowski's distribution as the ideal time of arrival distribution of collapsing states at the origin.

\section{Discussion}
From a purely technical point of view, we have shown that Kijowski's 
(continuum) time-of-arrival 
distribution for free, unconfined quantum particles can be obtained 
as the limit of the 
(discrete) distribution that results from quantizing the time of arrival 
in a finite box, taking into account the 
eigenvalue spacing 
in the transition from sums to integrals. Note that  
the TOA operator is not self-adjoint in the unconfined 
case, but it is self-adjoint in the box. There are several possible 
readings of this fact. We will point out some of them,
with the aim  
of opening a public discussion on the fundamental issues involved,
rather than to settle final answers and exhaustive 
explanations.   

For those following a traditional,
von Neumann's formulation of
quantum mechanics, the connection with a self-adjoint operator 
may be satisfactory since, within this framework, all observables must be  
associated with self-adjoint operators. 
One of the customary roles of self-adjointness is to assure the orthogonality 
of eigenfunctions so that, according to the projection postulate, 
repeated measurements would give the same result. It turns out however, that 
this idealization is not applicable to many observables and/or measurement 
operations. In particular, improper eigenfunctions for observables
with continuum spectrum 
(such as momentum, energy or position on the line)
are not normalizable,
and thus the collapse and repeated measurement idea cannot
be applied 
literally, even in principle, but only approximately.
It seems then that, as long as some ideal probability distribution 
may be computed unambiguously, there is no fundamental need to 
have a self-adjoint operator in these cases. 
Indeed, if one adopts the point of view that
observables are best 
formulated in terms of POVMs \cite{povm,Gia}, self-adjointness is not
essential 
to describe observables. The eigenfunction orthogonality may however be
seen anyway as a desirable, simplifying 
property,  whereby a part (component) of the initial state 
is responsible for a given outcome (eigenvalue) and not for any other.  
What the present paper shows is that the time of arrival eigenfunctions 
in the unbounded case differ,
appart from a normalization factor, only outside the large box or far
from the arrival point at the origin. The nonorthogonality in the continuum 
is thus caused by the distant eigenfunction behaviour with negligible 
overlap with the 
the initial wavepacket, and is thus physically irrelevant for computing the 
distribution.

Another problematic matter is the interpretation of the discrete nature of 
the time of arrival eigenvalues in the box. A discretization of time variables 
should not be surprising. For any system with discrete energies
or eigenmodes 
the corresponding 
time periods are also discrete. In the same vein, 
the TOA operator considered involves 
the discretized momentum operator in the denominator and therefore 
discrete eigenvalues. The problem thus 
is not in accepting the possibility of a discretization but in determining 
its operational meaning. 
``Measurements'', and ``observables'' are frequently highly idealized 
in quantum mechanics, to the point that all explicit reference to an
apparatus could disappear. This is useful, but may also let us 
without important physical references for its operational interpretation.   
Whereas an operational interpretation exists for Kijowski 
distribution in the continuum, a direct operational interpretation of
the discrete times 
of the confined case is still missing and is one of the challenges
for future research.      
In any case, the discrete-continuum smooth transition found here 
may be a useful tool to generate new theories of, say, first time-of-arrival 
with interacting potentials,
that can be later translated to the continuum and 
operationally interpreted or compared with existing operational 
proposals \cite{heg}.

\begin{acknowledgments}

This work was supported by 
``Ministerio de Ciencia y Tecnolog\'\i a-FEDER''
(BFM2003-01003), and
UPV-EHU (Grant 00039.310-15968/2004). 
The work of EAG is supported by the University of the Philippines through the
U.P. Creative and Research Scholarship Program.
ILE acknowledges partial support by the Spanish
Science Ministry under Grant FPA2002-02037.
\end{acknowledgments}


\begin{thebibliography}{00}

\bibitem{pauli} W. Pauli, {\em Hanbuch der Physik} vol V/1 ed. S Flugge 
(Springer-Verlag, Berlin,1926) 60.

\bibitem{oth} Y. Aharonov and D. Bohm,  {\it Phys. Rev.} {\bf122}, 1649 (1961); K. Kraus, \textsl{Zeitschrift fur Physik} \textbf{188}, 374 (1965); D.M. Rosenbaum, {\it J. Math. Phys.} {\bf 10}, 1127 (1969). M. Jammer {\it The philosophy of quatum mechanics} Wiley (1974); A. Peres, \textsl{Am. J. Phys.} \textbf{7} 552 (1980); V.S. Olkhovsky et al, {\it Nuovo Cimento} {\bf 22}, 263 (1974); P.R. Holland \textsl{The Quantum Theory of Motion} (1993). Press 	Syndicate, Cambridge University Press. Ph. Blanchard and A. Jadczyk, {\it Helv. Phys. Acta} {\bf 69} 613-635 (1996); J. Leon, {\it J. Phys. A} {\bf 30}, 4791 (1997).

\bibitem{povm} P. Busch, M. Grabowski and P. Lahti {\em Operational Quantum Physics} (Springer 1995); P. Busch, Found. Phys. {\bf 20}, 1, 33 (1990); M.D. Srinivas \& R. Vijayalakshmi, Pramana {\bf 16}, 173 (1981); M. Toller, Phys. Rev. A {\bf 59}, 960 (1999); P. Busch et.al. {\it An. Phys.} {\bf 237}, 1 (1995); P. Busch et. al., Phys. Let. A {\bf 191}, 357 (1994); H. Atmanspacher and A. Amann, {\it Int. J. Theo. Phys.} 629 (1998).

\bibitem{time} J. G. Muga  and C.R. Leavens, Phys. Rep. {\bf 338}, 353 (2000), and references therein.

\bibitem{gen} J.G. Muga, R. Sala Mayato, I.L. Egusquiza 
eds. {\it Time in Quantum Mechanics} (Springer 2002).

\bibitem{Kijowski74} J. Kijowski, Rep. Math. Phys. {\bf 6}, 362 (1974). 

\bibitem{All} G. R. Allcock, {Ann. Phys.} \textbf{53}, 253 (1969).

\bibitem{Gia} R. Giannitrapani, Int. J. Theor. Phys. {\bf 36}, 1575 (1997).

\bibitem{galapon} E.A. Galapon, Proc. R. Soc. Lond. A {\bf 487}, 1 (2002), quant-ph/9908033; E. A. Galapon {\em Proc. R. Soc. Lond. A} {\bf 458}, 451 (2002), quant-ph/0111061; E.A. Galapon quant-ph/0303106 . 

\bibitem{GCB04} E.A. Galapon, R. Caballar, and R.T. Bahague,
Phys. Rev. Let. {\bf 93}, 180406 (2004), quant-ph/0504174; E.A. Galapon quant-ph/0001062.

\bibitem{MLP98} J.G. Muga, C.R. Leavens, \& J.P. Palao, 
Phys. Rev. A {\bf 58} 4336, (1998).

\bibitem{BSPME00} A.D. Baute, R. Sala Mayato, J.P. Palao, J.G. Muga,
I.L. Egusquiza, 
Phys. Rev. A {\bf 61}, 022118 (2000).

\bibitem{LJPU00} J. Leon, J. Julve, P. Pitanga, and F.J. Urries
{\em Phys. Rev. A} {\bf 61} 062101 (2000).

\bibitem{BEM01b} A. D. Baute, I. L. Egusquiza and J. G. Muga, 
Phys. Rev. A {\bf 64}
012501 (2001).

\bibitem{BEM02} A.D. Baute, I.L. Egusquiza and J.G. Muga
{\em Phys. Rev. A} {\bf 65} 032114 (2002).

\bibitem{DEHM03} J.A. Damborenea, I.L. Egusquiza, G.C. Hegerfeldt and J.G. Muga 
{\em Phys. Rev. A} {\bf 66} 052104 (2002).

\bibitem{HSM03} G.C. Hegerfeldt, D. Seidel, and J.G. Muga
{\em Phys. Rev. A} {\bf 68} 022111 (2003).

\bibitem{EM00}
I.L. Egusquiza and J.G. Muga, Phys. Rev \textbf{A61}, 012104 (2000).
\newblock See also erratum, Phys. Rev. A \textbf{61} (2000) 059901(E).

\bibitem{BM94} A. J. Bracken and G. F. Melloy, J. Phys. A: Math. Gen {\bf 27}, 
2197 (1994). 

\bibitem{heg} G.C. Hegerfeldt, D. Seidel, J.G. Muga and
B. Navarro, Phys. Rev. A {\bf 70}, 012110 (2004).

\bibitem{grot} N. Grot, C. Rovelli, and R. Tate, {\it Phys.Rev. A} {\bf 54} 467 (1996).

\end{thebibliography}
\end{document}